\documentclass{aa}
\usepackage{amsmath,graphicx,amssymb}
\usepackage[varg]{txfonts}
\usepackage{natbib}
\newcommand{\zav}[1]{\left(#1\right)}
\newcommand{\hzav}[1]{\left[#1\right]}

\newlength\staretab

\newcommand\kms{\ensuremath{\text{km}\,\text{s}^{-1}}}

\newcommand{\teff}{\ensuremath{T_\text{eff}}}
\newcommand{\logg}{\ensuremath{\log (g/1\,\text{cm}\,\text{s}^{-2})}}

\newcommand{\lgep}[1]{\log\varepsilon_\text{#1}}

\begin{document}

\title{The nature of medium-period variables on the extreme horizontal branch
I. X-shooter study of variable stars in the globular cluster
$\omega$~Cen\thanks{Based on observations collected at the European Southern
Observatory, Paranal, Chile (ESO programme 108.224V).}}

\author{J.~Krti\v cka\inst{1} \and I.~Krti\v ckov\'a\inst{1} \and
        C.~Moni Bidin\inst{2} \and M.~Kajan\inst{1} \and 
        S.~Zaggia\inst{3} \and L. Monaco\inst{4} \and
        J.~Jan\'\i k\inst{1} \and
        Z.~Mikul\'a\v sek\inst{1} \and E.~Paunzen\inst{1}}

\institute{Department of Theoretical Physics and Astrophysics, Faculty of
           Science, Masaryk University, Kotl\'a\v rsk\' a 2, Brno,
           Czech Republic \and
           Instituto de Astronom\'\i a, Universidad Cat\'olica del Norte,
           Av.~Angamos, 0610 Antofagasta, Chile \and
           INAF – Padova Observatory, Vicolo dell’Osservatorio 5, 35122 Padova,
           Italy \and
           Universidad Andres Bello, Facultad de Ciencias Exactas,
           Departamento de Ciencias Físicas -- Instituto de Astrofisica,
           Autopista Concepción-Talcahuano, 7100, Talcahuano, Chile
}

\date{Received}

\abstract{A fraction of the extreme horizontal branch stars of globular clusters exhibit
a periodic light variability that has been attributed to rotational modulation caused
by surface spots. These spots are believed to be connected to inhomogeneous
surface distribution of elements. However, the presence of such spots has not
been tested against spectroscopic data. We analyzed the phase-resolved ESO
X-shooter
spectroscopy of three extreme horizontal branch stars that are members of the
globular cluster $\omega$ Cen and  also display periodic light variations. The aim of our
study is to understand the nature of the light variability of these stars and
to test whether the spots can reproduce the observed variability. Our spectroscopic
analysis of these stars did not detect any phase-locked abundance variations
that are able to reproduce the light variability. Instead, we revealed the phase variability
of effective temperature and surface gravity. In particular, the stars show the
highest temperature around the light maximum. This points to pulsations as a
possible cause of the observed spectroscopic and photometric variations. However,
such an interpretation is in a strong conflict with Ritter's law, which{
relates the pulsational period to the mean stellar density}. The location of the
$\omega$ Cen variable extreme horizontal branch stars in HR diagram corresponds
to an extension of PG~1716 stars toward lower temperatures or blue, low-gravity,  large-amplitude
pulsators toward lower luminosities, albeit with much longer periods. Other
models of light variability, namely, related to temperature spots, should  also
be  tested further. The estimated masses of these stars in the range
of $0.2-0.3\,M_\odot$ are too low for helium-burning objects.}


\authorrunning{J.~Krti\v{c}ka et al.}
\titlerunning{Nature of medium-period variables on extreme horizontal branch I.}
\maketitle

\section{Introduction}

A class of main sequence stars, called chemically peculiar stars, shows an
unusual type of light variability connected to the presence of surface spots
\citep{hummer,mobstersi}. These spots appear as a result of elemental diffusion,
whereby certain elements diffuse upwards under the influence of radiative force, while
others sink down as a result of gravitational pull
\citep{vimiri,ales3d,newdeal}. Moderated by the magnetic field (and also
by some additional processes, perhaps), surface inhomogeneities appear
\citep{kochrjab,jagelka}. The inhomogeneous surface elemental distribution,
together with the stellar rotation, leads to periodic spectrum variability. Additionally, the flux redistribution that is due to bound-bound (line) and bound-free
(ionization) processes modulated by the stellar rotation causes photometric
variability \citep{peter,ministr,molnar,lanko}. Based on abundance maps from
spectroscopy, we see that this effect is able to reproduce the observed rotational light variability of
chemically peculiar stars \citep{prvalis,myacen}.

Besides the radiative diffusion, chemically peculiar stars show other very
interesting phenomena, including magnetospheric radio emission
\citep{letovice,dasosm}, trapping of matter in circumstellar magnetosphere
\citep{labor,towo}, magnetic braking \citep{town}, and torsional variations
\citep{mikbra2}. However, up to now, such phenomena seems to be strictly
confined to classical chemically peculiar stars, which inhabit a relatively wide
strip on the main sequence with effective temperatures of about $7000-25\,000\,$K.

Therefore, it is highly desirable to search for other types of stars that show
similar phenomena. The most promising candidates are stars that have signatures
of radiative diffusion in their surface abundances, such as hot horizontal
branch stars \citep{un,miriri} and hot white dwarfs \citep{chadif,unbu}. Indeed,
variations of helium to hydrogen number density ratio have been found on the
surface of white dwarfs \citep{hebheprom,janus} and some extremely hot white
dwarfs even show signatures of corotating magnetospheres \citep{baf} and spots
\citep{hotvar}.

The phenomena connected with chemically peculiar stars can be most easily traced
by periodic photometric light variability. However, while there are some
signatures of chemical spots in white dwarfs all along their cooling track
\citep{dupchaven,kilic,baf}, the search for light variability in field
horizontal branch stars with $T_\text{eff}<11\,000\,$K has turned out to be
unsuccessful \citep{pahori}. The prospect of rotationally variable hot subdwarfs
was further marred by the discovery of a handful of hot subdwarfs which, despite
their detectable surface magnetic fields \citep{dormag}, still do not show any light
variability \citep{trimagpod}.

This perspective changed with the detection of possible rotationally variable hot horizontal
branch stars of globular clusters by \citet{mombible}. However, the presence of
abundance spots was anticipated from photometry without any support from
spectroscopy. Therefore, we started an observational campaign aiming at
detection of abundance spots on these stars and understanding this type of
variability overall. Here, we present the results derived for members of
$\omega$~Cen (NGC~5139).

\section{Observations and their analysis}

We obtained the spectra of supposed rotational variables in NGC~5139 within the
European Southern Observatory (ESO) proposal 108.224V. The spectra were acquired
with the X-shooter spectrograph \citep{xshooter} mounted on the 8.2m Melipal
(UT3) telescope and these observations are summarized in Table~\ref{pozor}. The
spectra were obtained with the UVB and VIS arms providing an average spectral
resolution ($R=\lambda/\Delta\lambda$) of 5400 and 6500, respectively. Although
medium-resolution spectrograph is not an ideal instrument for abundance
analysis, the abundance determination is typically based on multiple strong
lines of given elements. This mitigates the disadvantages of the
medium-resolution spectra and enables us to estimate reliable abundances
\citep[e.g.,][]{kawnltt,gvarpol}. In turn, the use  of a medium-resolution
spectrograph implies a lower number of elements that can be studied and also worsens
the precision with respect to the abundance determinations in cases of spectral blends. We extracted
the calibrated spectra from the ESO archive. The radial velocity was determined
by means of a cross-correlation using the theoretical spectrum as a template
\citep{zvezimi} and the spectra were shifted to the rest frame.

\begin{table}[t]
\caption{Spectra used for the analysis.}
\centering
\label{pozor}
\begin{tabular}{cccc}
\hline
Spectrum (prefix) & JD$-2\,400\,000$ & Phase & S/N\\
\hline
\multicolumn{4}{c}{vEHB-2, $P=7.82858823\,$d}\\
\multicolumn{4}{c}{$\alpha=$ 13h 26m 22.572s,
$\delta=-47^\circ$ $30'$ $52.786''$}\\
XS\_SFLX\_3060818 & 59612.82872 & 0.014 & 36\\
XS\_SFLX\_3060821 & 59653.61422 & 0.224 & 18\\
XS\_SFLX\_3060824 & 59600.77245 & 0.474 & 21\\
XS\_SFLX\_3060830 & 59627.79067 & 0.925 & 36\\\hline
\multicolumn{4}{c}{vEHB-3, $P=5.16509016\,$d}\\
\multicolumn{4}{c}{$\alpha=$ 13h 26m 21.922s,
$\delta= -47^\circ$ $26'$ $05.753''$}\\
XS\_SFLX\_3073156 & 59597.79411 & 0.276 & 17\\
XS\_SFLX\_3073156 & 59623.74778 & 0.301 & 35\\
XS\_SFLX\_3073159 & 59603.79086 & 0.437 & 36\\
XS\_SFLX\_3073165 & 59589.77704 & 0.724 & 28\\
XS\_SFLX\_3073168 & 59621.83821 & 0.931 & 41\\
XS\_SFLX\_3073168 & 59627.74979 & 0.076 & 34\\\hline
\multicolumn{4}{c}{vEHB-7, $P=1.78352993\,$d}\\
\multicolumn{4}{c}{$\alpha=$  13h 27m 17.454s,
$\delta=  -47^\circ$ $27'$ $49.059''$}\\
XS\_SFLX\_3060944 & 59667.56405 & 0.404 & 38\\
XS\_SFLX\_3060950 & 59651.84211 & 0.589 & 45\\
XS\_SFLX\_3060956 & 59600.82010 & 0.982 & 30\\
\hline
\end{tabular}
\tablefoot{Photometric periods and J2000 coordinates determined by
\citet{mombible}. The phases were calculated for arbitrary
JD$_0=2\,458\,031.346$. S/N is a median value.}
\end{table}

\begin{table}[t]
\caption{List of wavelengths (in \AA) of the strongest lines used for abundance
determinations.}
\label{hvezdadcar}
\begin{tabular}{ll}
\hline
\ion{He}{i}  & 3820, 4009, 4024, 4026, 4144, 4388, 4471, 4713, \\
             & 4922, 5016\\
\ion{C}{ii}  & 3876, 3919, 3921, 4267 \\
\ion{N}{ii}  & 3995, 4035, 4041, 4447, 4631, 5001 \\
\ion{O}{ii}  & 3954, 4396, 4415 \\
\ion{Mg}{ii} & 4481 \\
\ion{Al}{iii}& 4513, 4529 \\
\ion{Si}{ii} & 3856, 3863 \\
\ion{Si}{iii}& 3807, 4553, 4568, 4575 \\
\ion{Ca}{ii} & 3934 \\
\ion{Fe}{iii}& 3954, 4035, 4138, 4139, 4165, 4273, 4286, 4297, \\
             & 4372, 4396, 4420, 4431, 5127, 5194 \\
\hline
\end{tabular}
\end{table}

The stellar parameters were determined using the simplex minimization \citep{kobr}
in three steps. First, we determined the effective temperature, $T_\text{eff}$,
and the surface gravity, $\log g$, by fitting each of the observed spectra with
spectra derived from the BSTAR2006 grid of NLTE\footnote{The NLTE models allow
for departures from the local thermodynamic equilibrium (LTE) due to radiative
processes.} plane parallel model atmospheres with $Z/Z_\odot=0.1$
\citep{bstar2006}. For the present purpose, the grid was extended for models with
$\log g=5$. The random errors of $T_\text{eff}$ and $\log g$ for individual
observations were determined by fitting a large set of artificial spectra
derived from observed spectra by the addition of random noise with a Gaussian
distribution. The dispersion of noise was determined by the signal-to-noise ratio
(Table~\ref{pozor}).

We then estimated surface abundances using the model atmosphere from the grid
located closest to the mean of derived parameters. The abundance determination
was repeated once more using NLTE plane parallel model atmospheres calculated
with TLUSTY200 \citep{ostar2003,bstar2006} for parameters derived in the previous
steps. To determine the abundances, we matched the synthetic spectra calculated by
SYNSPEC49 code with observed spectra. The random errors of abundances for
individual observations were also determined by fitting of artificial spectra
derived by adding random noise to the observed spectra. For elements whose
abundances were not derived from spectra, we assumed a typical $\omega$~Cen
abundance $\log(Z/Z_\odot)=-1.5$ \citep{momega,moniomega}. The spectral lines used
for the abundance analysis are listed in Table~\ref{hvezdadcar}. The final
parameters averaged over individual spectra are given in Table~\ref{hvezpar}.
The derived individual elemental abundances are expressed relative to hydrogen
$\varepsilon_\text{el}=\log(n_\text{el}/n_\text{H})$. Random errors given in
Table~\ref{hvezpar} were estimated from parameters derived from the fits of
individual spectra.

\begin{table*}[t]
\caption{Derived parameters of the studied stars.}
\label{hvezpar}
\centering
\begin{tabular}{lccccc}
\hline
Parameter       & vEHB-2           & vEHB-3 & vEHB-7 & Sun \\
\hline
\teff\ [K]      & $24\,900\pm1200$ & $21\,000\pm1200$ & $21\,200\pm600$\\
\logg           & $4.81\pm0.12$    & $4.64\pm0.04$  & $4.70\pm0.04$\\
$R$ [$R_\odot$] & $0.34\pm0.05$    & $0.39\pm0.06$  & $0.33\pm0.03$ \\
$M$ [$M_\odot$] & $0.27\pm0.11$    & $0.25\pm0.08$  & $0.20\pm0.04$ \\
$L$ [$L_\odot$] & $40\pm7$         & $27\pm5$       & $20\pm3$ \\
$\lgep{He}$     & $-3.0\pm0.2$     & $-3.3\pm0.5$   & $-2.9\pm0.2$ & $-1.07$ \\
$\lgep{C}$      & $-4.5\pm0.1$     & $-4.7\pm0.2$   & $-4.9\pm0.1$ & $-3.57$ \\
$\lgep{N}$      & $-4.6\pm0.1$     & $-4.4\pm0.1$   &              & $-4.17$ \\
$\lgep{O}$      & $-4.6\pm0.4$     & $-4.2\pm0.2$   &              & $-3.31$ \\
$\lgep{Mg}$     &                  & $-5.3\pm0.1$   & $-5.9\pm0.1$ & $-4.40$ \\
$\lgep{Al}$     & $-6.0\pm0.1$     &                &              & $-5.55$ \\
$\lgep{Si}$     & $-5.1\pm0.1$     & $-5.0\pm0.1$   & $-4.6\pm0.1$ & $-4.49$ \\
$\lgep{Ca}$     &                  & $-6.0\pm0.1$   & $-5.4\pm0.8$ & $-5.66$ \\
$\lgep{Fe}$     & $-3.8\pm0.1$     & $-4.2\pm0.2$   & &              $-4.50$ \\
$v_\text{rad}$ [\kms] & $195\pm7$  & $231\pm4$      & $227\pm3$ \\
\hline
\end{tabular}\\
\tablefoot{Solar abundances were taken from \citet{asp09}. Blank items denote
values that were not determined.}
\end{table*}

\section{Analysis of individual stars}

\subsection{Star vEHB-2}
\newcommand\hvezda{vEHB-2}

\begin{figure}[t]
\includegraphics[width=0.5\textwidth]{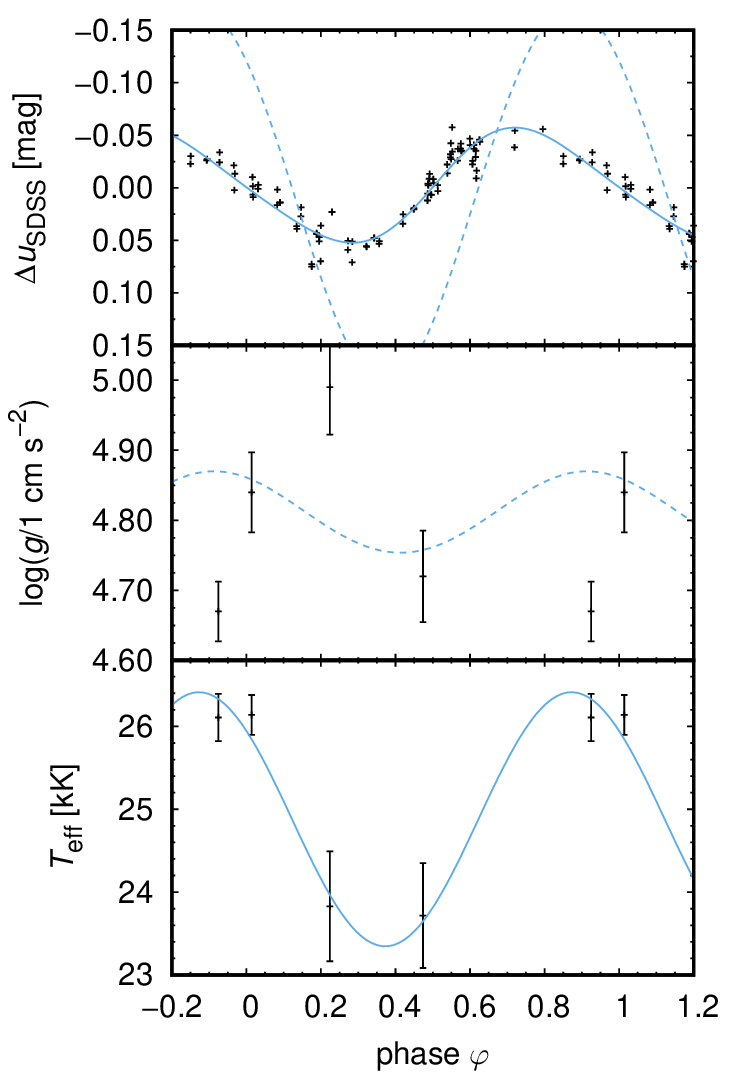}
\caption{Phase variations of \hvezda. {\em Upper panel}: Observed light
variations from \citet{mombible}. Dashed blue line denotes predictions deduced
purely from temperature variations, while solid line denotes a fit with
additional sinusoidal radius variations. {\em Middle panel}: Surface gravity
variations. Dashed blue line denotes surface gravity determined from the radius
variations. {\em Lower panel}: Effective temperature variations. Solid blue
line denotes sinusoidal fit. Part of the variations for $\varphi<0$ and
$\varphi>1$ are repeated for better visibility.}
\label{ehb2prom}
\end{figure}

Our analysis of the spectra for the star \hvezda\ (listed in Table~\ref{pozor}) revealed
periodic changes in surface gravity and effective temperature (see
Fig.~\ref{ehb2prom}). To test their presence and any possible correlations, we fixed
either the surface gravity or effective temperature and repeated the fit to
determine the missing parameter. The test revealed a similar variability of
the effective temperature and surface gravity as derived from the fit of both
parameters and has not shown any significant change of the derived parameters.
Neither one of the parameters determined from individual spectra with added random
noise showed any strong correlations. Thus, we conclude that the
detected variations of surface gravity and effective temperature are real. We
did not detect any strong phase variations of elemental abundances or radial
velocities (Sects.~\ref{kapabskvrn} and \ref{kapdvoj}).

Table~\ref{hvezpar} lists derived parameters of \hvezda\ averaged over the available
spectra. The abundances of many elements is slightly higher than a typical
$\omega$~Cen composition $\log(Z/Z_\odot)=-1.5$ \citep{momega,moniomega}. The
exceptions are helium, which is strongly underabundant as a result of
gravitational settling, and iron, whose overabundance can be interpreted as a
result of radiative diffusion \citep{unbuhb,miriri}.

%2012A&A...547A.109M 
With $V=17.249\,\text{mag}$ \citep{mombible}, $E(B-V) = 0.115 \pm
0.004\,\text{mag}$ \citep{moniomega}, the bolometric correction of \citet[see
also \citealt{spravnybyk}]{bckytka} $\text{BC} = -2.40\pm0.12\,\text{mag}$, and
distance modulus $(m - M)_0 = 13.75 \pm 0.13\,\text{mag}$ \citep{distom}, the
estimated luminosity is $L=40\pm7\,L_\odot$. With determined atmospheric
parameters this gives the stellar radius $0.34\pm0.05\,R_\odot$ and mass
$0.27\pm0.11\,M_\odot$. Derived effective temperature is slightly lower than the
estimate $28\,200 \pm 1600$\,K of \citet{momega}, while our results agree with
their surface gravity, $\log g=4.86 \pm 0.18,$ and helium abundance,
$\log\varepsilon_\text{He}=-3.20 \pm 0.16$.

\subsection{Star vEHB-3}
\renewcommand\hvezda{vEHB-3}

\begin{figure}[t]
\includegraphics[width=0.5\textwidth]{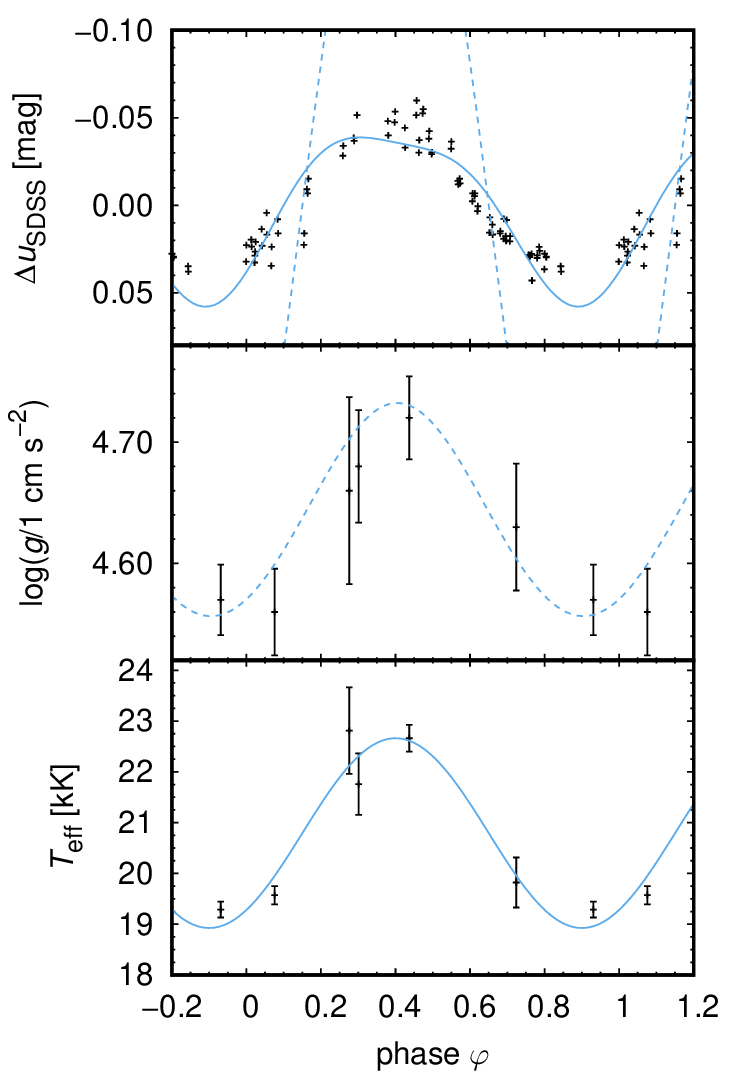}
\caption{Same as Fig.~\ref{ehb2prom}, but for \hvezda.}
\label{ehb3prom}
\end{figure}

The spectral analysis of \hvezda\ also revealed phase-locked variability of the
effective temperature and surface gravity (Fig.~\ref{ehb3prom}). The star is
hotter and shows higher surface gravity during the light maximum. The analysis
of individual spectra has not revealed any significant variations of the radial
velocity (Sect.~\ref{kapdvoj}).

A detailed inspection of spectra shows that the strength of helium lines is
variable. This can be most easily seen in \ion{He}{i} 4026\,\AA\ and 4471\,\AA\
%and \ion{Ca}{ii} 3934\,\AA\
lines (Sect.~\ref{kapcarprom}). In principle, such variability may also reflect
the temperature variations. To test this, for this star we determined abundances
for actual temperature and surface gravity derived from individual spectra and
not just for the mean values. Even with this modified approach the helium
abundance variations has not disappeared, showing that simple effective
temperature and gravity variations cannot reproduce the variability of helium
lines. We have not detected any strong variability of the line strengths of other
elements (Sect.~\ref{kapabskvrn}).

For \hvezda, \citet{mombible} gives $V=17.274\,\text{mag}$, while the mean reddening is
$E(B-V) = 0.115 \pm 0.004\,\text{mag}$ \citep{moniomega}, the bolometric
correction of \citet{bckytka} is $\text{BC} = -2.00\pm0.13\,\text{mag}$, and
the distance modulus is $(m - M)_0 = 13.75 \pm 0.13\,\text{mag}$ \citep{distom};
this results in the luminosity of $L=27\pm5\,L_\odot$. With the determined atmospheric
parameters, this gives a stellar radius of $0.39\pm0.06\,R_\odot$ and mass of
$0.25\pm0.08\,M_\odot$.

\subsection{Star vEHB-7}
\renewcommand\hvezda{vEHB-7}

\begin{figure}
\includegraphics[width=0.5\textwidth]{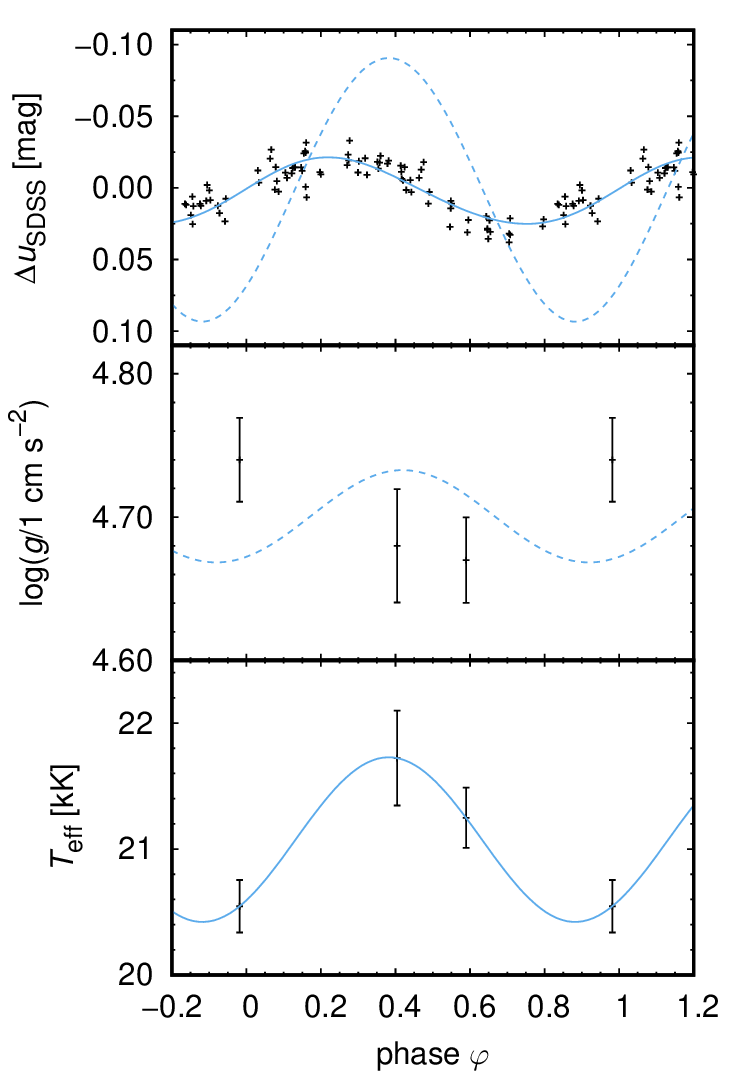}
\caption{Same as Fig.~\ref{ehb2prom}, but for \hvezda.}
\label{ehb7prom}
\end{figure}

In total, five spectra for \hvezda\ were obtained. However, one of them is of a
poor quality and an additional spectrum was marred by a wrong pointing.
Consequently, there are just three spectra left. Anyway, the analysis of these
spectra indicates presence of temperature variations (Fig.~\ref{ehb7prom}), with
temperature maximum appearing around the time of light maximum. The available
spectra do not show any strong variability of surface abundances nor the radial
velocities (Sects.~\ref{kapabskvrn} and \ref{kapdvoj}).

The $V$ magnitude of the star is $V=17.644\,\text{mag}$ \citep{mombible}, which
with the reddening $E(B-V) = 0.115 \pm 0.004\,\text{mag}$ \citep{moniomega}, the
bolometric correction of \citet{bckytka} $\text{BC} = -2.02\pm0.07\,\text{mag}$,
and distance modulus $(m - M)_0 = 13.75 \pm 0.13\,\text{mag}$ \citep{distom}
gives a luminosity of $L=20\pm3\,L_\odot$. With the atmospheric parameters
determined from spectroscopy, this gives a stellar radius of
$0.33\pm0.03\,R_\odot$ and mass of $0.20\pm0.04\,M_\odot$. The star was analyzed by
\citet{shotomcen}, who derived slightly higher effective temperature of $23800 \pm
800\,$K, surface gravity of $\log g=5.11 \pm 0.06 $, and mass  of $0.49 \pm
0.10\,M_\odot$.

\section{Significance of the detected variations}

Before discussing the implications of the detected variations for the mechanism
of the light variability of the stars, we  first need to clarify whether the detected
variations could be real. To this end, we used a random number generator
to create a population of stellar parameters with dispersions determined from the
uncertainty of each measurement in each phase. We compared the dispersion of the
derived artificial population with the dispersion of the derived data and determined
a fraction of the population that gives a higher dispersion than the data
determined from observation. If this fraction is high, then it is likely that
the derived variations are only sampling random noise.

For the effective temperature, the derived fraction is lower than $10^{-5}$ for
all three stars. The uncertainties of estimated effective temperatures should be
a factor of three higher to reach a fraction of $0.01$ for vEHB-7 -- and even
higher for the remaining stars. Thus, we conclude that the detected
variations of the effective temperature are very likely to be real for all the
stars studied here.

The same is true for the variations of the surface gravity, where the
uncertainties should by a factor of $1.6$ higher to reach a fraction of $0.01$
for vEHB-7. From this, we conclude that also the variations of the surface
gravity are very likely real in vEHB-2 and vEHB-3, with a small chance that the
gravity variations in vEHB-7 are random.

\section{Nature of the light variations}

\subsection{Pulsations}

We detected a variability among the effective temperature and surface gravity phased
with photometric variations in all studied stars
(Figs.~\ref{ehb2prom}--\ref{ehb7prom}). The effective temperature and surface
gravity are typically the highest during the maximum of the light variability.
In the absence of any strong radial velocity variations (Sect.~\ref{kapdvoj}), such
changes in the stellar parameters can be most naturally interpreted as resulting from the
pulsations \citep[e.g.,][]{wooj,foskol,vasil}.

To test the pulsational origin of the light variability, we calculated the
synthetic light curves and compared it with observed light variability. As a
first step, we used the fluxes from the BSTAR2006 database calculated for
$Z=0.1Z_\odot$ and $\log g=4.75$, convolved them with the response function of
$u_\text{SDSS}$, and fitted them as a function of the effective temperature,
deriving:
\begin{multline}
\label{usdss}
-2.5\log\zav{\frac{H(u_\text{SDSS})}
{1\,\text{erg}\,\text{s}^{-1}\,\text{cm}^{-2}}}=\\
-19.29-0.275\zav{\frac{\teff}{10^3\,\text{K}}}+
0.0032\zav{\frac{\teff}{10^3\,\text{K}}}^2.
\end{multline}
The fit is valid between $\teff=15-30\,$kK. We fit the observational phase
variations of the effective temperature by a simple sinusoidal (plotted in
Figs.~\ref{ehb2prom}--\ref{ehb7prom}) and used these variations to predict the
light variations (dashed curve in the upper plots of
Figs.~\ref{ehb2prom}--\ref{ehb7prom}). The prediction assumes that the
temperature is the same across the stellar surface, corresponding to the radial
pulsations.
The resulting light variations have
always higher amplitude than the observed light curve, but this can be
attributed to radius variations. We searched for such sinusoidal radius
variations that would allow us to reproduce the observed light variations. It
turns out that radius variations with amplitudes of about few percent and phase-shifted by nearly half period from temperature variations are fully able to reproduce
the observed light variations (solid line in the upper panels of
Figs.~\ref{ehb2prom}--\ref{ehb7prom}).  Assuming that the pulsating atmosphere
is roughly in hydrostatic equilibrium, the effective surface gravity varies due
to a change in radius and as a result of inertial force. This is plotted using the
dashed curve in the middle plot of Figs.~\ref{ehb2prom}--\ref{ehb7prom}. The
resulting amplitude of the surface gravity variations is always comparable to
the observed variations, albeit the curves  are in good agreement only for vEHB-3.

The fact that the resulting phase variations of surface gravity do not fully agree
with observations is understandable for several reasons. The spectroscopy was
obtained just in few phases, which makes the effective temperature phase curve
rather uncertain. Moreover, the width of the line profiles is affected by the
electron number density and not directly by the surface gravity. The dependence
of the line profiles on gravity stems from the hydrostatic equilibrium equation.
However, the equation of hydrostatic equilibrium can be violated in pulsating
stars, especially in the presence of shocks \citep{jefhydropul}. Additionally,
the effective temperature determined from spectroscopy may not correspond to
temperature of radiation emerging from the continuum formation region. Finally,
{contrary to our  assumption,}
the stars may experience non-radial pulsations, further complicating the analysis.

\begin{figure}
\includegraphics[width=0.5\textwidth]{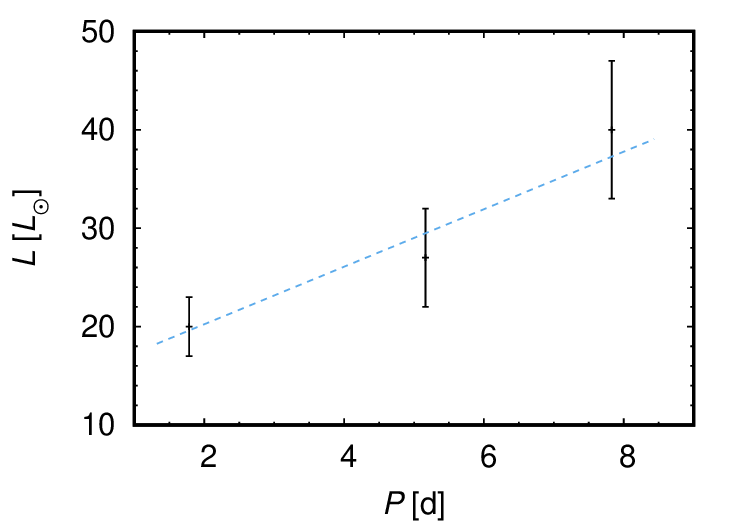}
\caption{Period-luminosity relationship for studied stars. Dashed line
corresponds to the linear fit.}
\label{perlum}
\end{figure}

Pulsating stars often show relation between period and luminosity
\citep[e.g.,][]{labut,fridmanova,moperlum} which stems from the dependence of
pulsational period on mean stellar density or sound wave crossing time. It is
worthy to notice that in Table~\ref{hvezpar} the more luminous stars have longer
periods. On average, the period-luminosity relationship can be expressed as
(see Fig.~\ref{perlum}):
\begin{equation}
\zav{\frac{L}{L_\odot}}=2.9\zav{\pm0.7}\zav{\frac{P}{1\,\text{d}}}+14\zav{\pm3}.
\end{equation}
However, the analysis involves strong selection effect, because we have focused on
brightest stars from the \citet{mombible} sample.

The pulsational hypothesis can be further tested using ultraviolet photometric
variations \citep[e.g.,][]{sovickaii}, which should correspond to optical
variations. The amplitude of the radial velocity variations due to proposed
pulsational motion is of the order $0.1\,\kms$. Therefore, the presence of
pulsations can be also tested using precise radial velocity measurements.

However, the interpretation of observed light variations in terms of pulsations
poses a challenge for pulsational theory. Field hot subdwarfs typically pulsate
with frequencies that are one to two orders of magnitude higher than found here
\citep{osten,olovo,bartes}. This stems from Ritter's law \citep{ritter},
which predicts that the period of pulsations is inversely proportional to the
square root of the mean stellar density. As a result, tenuous cool giants and
supergiants pulsate with periods of the order of hundreds of days
\citep{cervobrpul}. On the other hand, the p-modes of relatively high-density
hot subdwarfs are predicted to have periods of the order of hundreds of seconds
\citep{guoslup}. With typical pulsational constants \citep{les,gaucovka},
Ritter's law gives a period of the order of hundredths of a day for studied
stars, which is three orders of magnitude lower than the period of variability
of studied stars. The beating of two close periods could lead to
variability with longer period, but it remains unclear how the short periods could
be damped in surface regions.
The g-modes may have longer periods \citep{milbezablesk} and
would thus serve as  better candidates for explaining the observed periodic light
variability.

The period of g-mode pulsations depends on the buoyancy oscillation travel
time across the corresponding resonance cavity \citep{gardlouhytess}. The
related Brunt–V\"ais\"al\"a frequency approaches zero when the radiative
temperature gradient is close to the adiabatic gradient. Hot stars possess an iron
convective zone, which disappears for low iron abundances \citep{jermin}.
However, the studied stars show relatively high iron abundance as a result of
radiative diffusion (Table~\ref{hvezpar}). Therefore, it is possible that
interplay of the radiative diffusion and proximity to the convection instability
may lead to the appearance of medium-period pulsations.

The pulsations may not necessarily be driven by classical $\kappa$-mechanism.
The location of studied stars in $\log g - T_\text{eff}$ diagram corresponds to
stars experiencing helium subflashes before the helium-core burning phase
\citep{batika}. Such stars are predicted to have pulsations driven by the
$\epsilon$-mechanism.

If the light variations are indeed due to pulsations, then the stars could be
analogues of other pulsating subdwarfs, as the EC~14026 stars \citep{ececko} and
PG~1159 stars \citep[GW Vir stars,][]{corsika}, however, with much longer
periods. Taking into account the derived stellar parameters, the location of the
variables from $\omega$~Cen in HR diagram corresponds to the extension of
PG~1716 stars \citep{zelpulnov} toward lower effective temperatures. Stellar
parameters of studied stars are also close to the blue large-amplitude pulsators
\citep{blap}, which are somehow more luminous and slightly hotter. The search
for pulsations in corresponding cluster stars was, to our knowledge, not
successful \citep{necti}; surprisingly, only significantly hotter pulsating
stars were detected on the horizontal branch \citep{randal,browpul}. The studied
stars are located in area of Hertzsprung-Russell diagram (HRD), where pulsations
resulting from the $\kappa$-mechanism on the iron-bump opacity can be expected;
however, with significantly shorter periods \citep{ostrofont,jefpodpul,jefsapul}.
The pulsational instability appears at high iron abundances, which are also
detected in studied stars.

Pulsating subdwarfs typically evince non-radial pulsations \citep{corsika}, for
which low amplitudes of photometric variations are expected. Still,
\citet{kurad} detected a new class of variable stars corresponding to
blue, high-gravity, large-amplitude pulsators that are pulsating radially with
amplitudes that are comparable to the stars studied here.

\subsection{Abundance spots}
\label{kapabskvrn}

\begin{figure}
\centering
\includegraphics[width=0.5\textwidth]{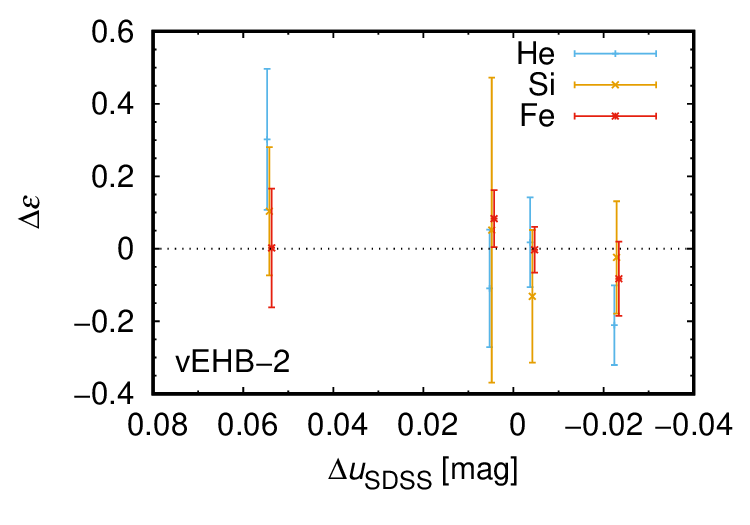}
\includegraphics[width=0.5\textwidth]{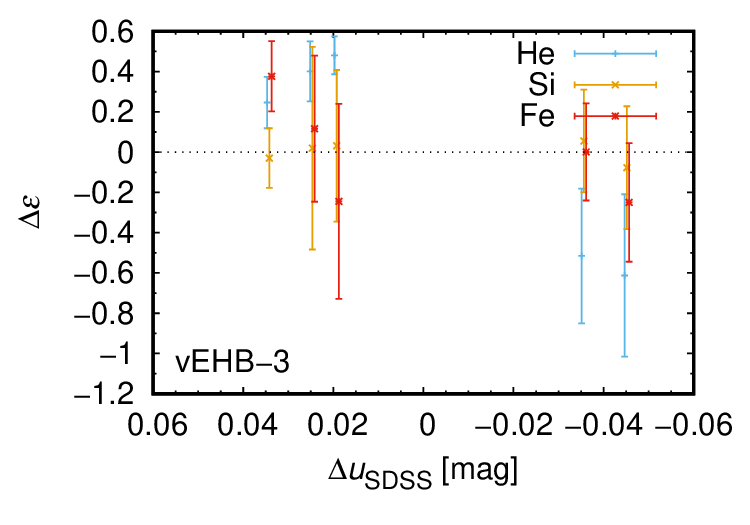}
\includegraphics[width=0.5\textwidth]{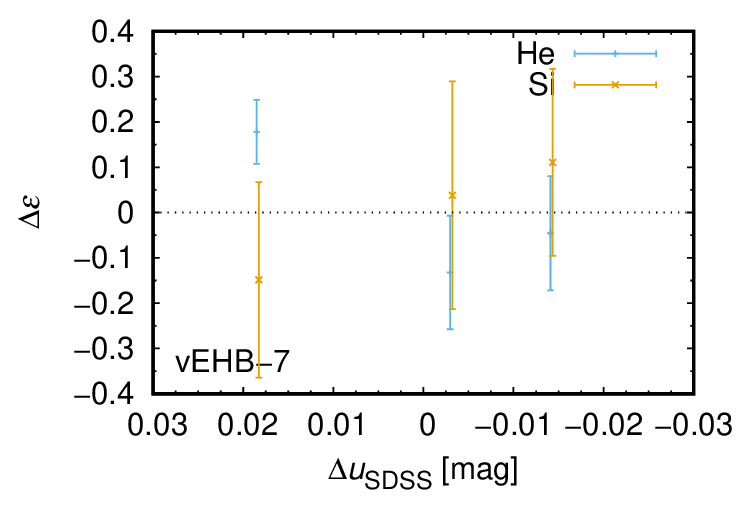}
\caption{Difference between abundances of selected elements derived from
individual spectra and a mean abundance. Plotted as a function of relative
magnitudes for individual stars. Elements plotted in the graph typically
contribute most significantly to the light variations at studied effective
temperatures \citep{mysigorie,myacen}. The individual points were shifted slightly
horizontally  to avoid overlapping.}
\label{fazsloz}
\end{figure}

As one of the possible mechanisms behind the detected light variability,
\citet{mombible} suggested the rotational flux modulation due to abundance
spots. Any light variability modulated by rotation requires that the rotational
velocity determined from the period of variability and stellar radius should be
higher than the rotational velocity projection, $v_\text{rot}\sin i$, determined from
spectroscopy. However, the spectroscopy provides only a very loose constraint on
the rotational velocities of individual stars, $v_\text{rot}\sin i<50\,\kms$. As
a result, with the stellar radii (from Table~\ref{hvezpar}) and photometric
periods (listed in Table~\ref{pozor}) determined by \citet[]{mombible}, the
rotational modulation of photometric variability cannot be ruled out. Therefore,
the test of abundance spots requires a more elaborate approach.

In principle, determination of light curves due to abundance spots from the
observed spectroscopy is a straightforward procedure. The inverse method of
Doppler imaging is used to determine surface abundance maps
\citep[e.g.,][]{kofidra}. From the derived abundance maps, the light curves can
be simulated using model atmospheres synthetic spectra \citep[e.g.,][]{myacen}.
However, the Doppler imaging requires relatively large number of high-resolution
and high signal-to-noise-ratio (S/N) spectra. With the current instrumentation, this is
beyond the reach of even 8-m class telescopes. Therefore, another method has to
be used to test the presence of surface spots.

For faint stars, it is possible to estimate surface abundances as a function of
phase and simulate the light variability directly from derived abundances
\citep{esobtprmhvi}. However, the observations do not suggest the presence of
abundance spots on the surface of the stars. There is some scatter of abundances
derived from individual spectra, but the potential abundance variations are not
correlated with light variations. This can be seen from Fig.~\ref{fazsloz},
where we plot the abundances derived from individual spectra as a function of observed
magnitude (both values are plotted with respect to the mean). If the light
variations were due to the abundance variations, the plot should evince a positive
correlation between abundance and magnitudes \citep{prvalis,myacen}, but such
a correlation is missing. Moreover, the amplitude of abundance variations (which
is no more than about 0.1\,dex) should be one magnitude higher to cause observed
light variations \citep[c.f.,][]{mysigorie,myacen}. On top of that, the mean
abundance should be high enough to affect the emergent flux.

We additionally tested the abundance spot model of the light variability
using model atmosphere emergent fluxes. We calculated the model atmospheres with ten
times higher abundances of helium, silicon, and iron than those determined from
spectroscopy. This is an order of magnitude higher overabundance than
observations allow. We calculated the magnitude difference between the fluxes
corresponding to enhanced and observational abundances in the $u_\text{SDSS}$
band used by \citet{mombible}. This gives a theoretical upper limit of the
magnitude of the light variability. In the case of helium and silicon, the
derived amplitude of the light variability would be 0.002\,mag and 0.02\,mag,
which is significantly lower than the observed amplitude of the light
variability. The amplitude is higher only in the case of iron (0.4\,mag), but
even in this case the maximum iron abundance does not appear during the maximum
of the light curve (Fig.~\ref{fazsloz}). Moreover, the detected abundance
variations can be interpreted in terms of random fluctuations.

There is a possibility that the variations are caused by element(s) that do not
appear in the optical spectra. However, this is unlikely, because in classical
chemically peculiar stars the abundance variations are not confined just to a
single element \citep[e.g.,][]{ruskol,kofidra}. Consequently, we conclude that
derived abundance variations from individual spectra are most likely of
statistical origin. Therefore, the studied stars do not likely show light
variability due to surface spots similar to main sequence, chemically peculiar
stars.

The only element that varies with magnitude is helium (Fig.~\ref{fazsloz}), but
it shows opposite behavior than is required to explain the light variability due
flux redistribution. This means that the helium lines are observed to be
stronger during the light minimum. Moreover, the helium line profiles are
complex and we were unable to reasonably fit the observed helium lines using
synthetic spectra. Therefore, instead of spots, we suspect that they are formed
by intricate motions in the atmosphere during pulsations
(Sect.~\ref{kapcarprom}).

\subsection{Binary origin}
\label{kapdvoj}

\begin{figure}
\centering
\includegraphics[width=0.5\textwidth]{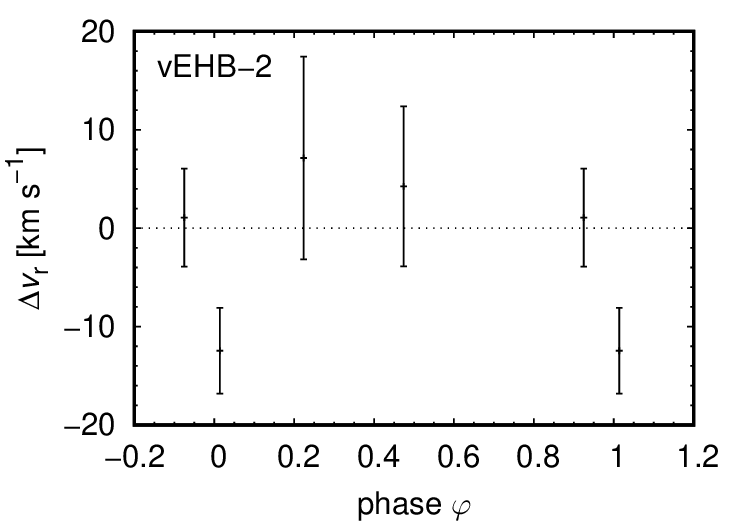}
\includegraphics[width=0.5\textwidth]{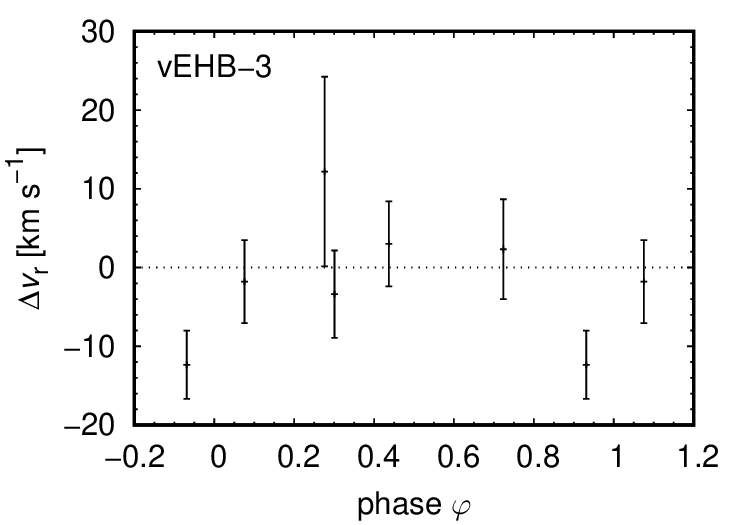}
\includegraphics[width=0.5\textwidth]{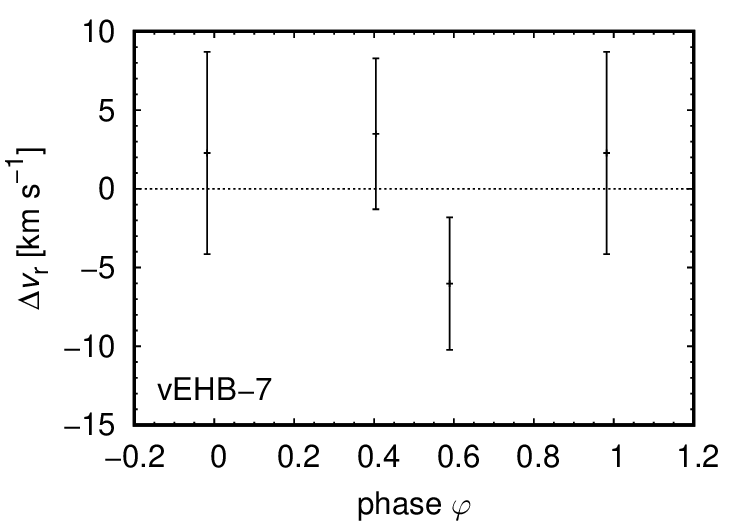}
\caption{Phase variations of radial velocity determined from individual spectra
with respect to the mean value. Plotted for individual stars. Parts of the
variations for $\varphi<0$ and $\varphi>1$ repeat for a better visibility.}
\label{vr}
\end{figure}

It may be possible that the observed light variations are due to binary
effects. In that case, there would be a number of combinations for the arrangement of the system. It is unlikely that the variations are due to the reflection effect on
a cooler companion, because in such cases, the system would look cooler during the
light maxima, which would contradict the observations. Moreover, the predicted amplitude
would be too low. Due to the absence of any strong radial velocity variations
(Fig.~\ref{vr}) and for evolutionary reasons, the cooler companion would be
less massive; from this (and the third Kepler law), the resulting binary separation would
be about $11\,R_\odot$ for vEHB-2 (and even lower for remaining stars). This
once more precludes the assumption of the red giant as a companion and leaves just enough space for a
low-mass main sequence star. Furthermore, we used our code calculating light curves
due to the reflection effect, which predicts that the radius of such a star should
be comparable to solar radius to cause observed light variations. Therefore, for
low-mass main sequence star, the light amplitude would be significantly lower
than observed.

The reflection due to a hotter companion is constrained by the absence of strong
companion lines in the optical spectra and by missing large radial velocity
variations (Fig.~\ref{vr}). This leaves two options, both involving hot
(possibly degenerated) companion. Either the companion has low mass (possibly implying a hot helium white dwarf) or the system involves high-mass companion on
an orbit with low inclination. In any case, given a typical mass of extreme
horizontal branch stars \citep{moni6752,moniomega} and maximum mass of white
dwarfs \citep{yoshchandra,nunchandra}, it is unlikely that the total mass of the
system exceeds $2\,M_\odot$. In this case the Kepler third law predicts an
orbital separation of $a=21\,R_\odot$ for vEHB-2. From the Saha-Boltzmann law,
the required temperature of irradiating body is
$T_\text{irr}=\sqrt{a/R_\text{irr}} \hzav{2\zav{T_2^4-T_1^4}}^{1/4}$, where
$R_\text{irr}$ is the radius of irradiating body and $T_2$ and $T_1$ are the
maximum and minimum temperatures of studied star. With a typical radius of a
white dwarf $R_\text{irr}=0.01\,R_\odot$ this gives
$T_\text{irr}=10^6\,\text{K}$, far exceeding the temperature of any white dwarf
\citep{milbe}. This estimate could be decreased assuming a lower mass of
irradiating body and excluding the detected radius variations, but it still
amounts to about 300\,kK for the star vEHB-7 with the shortest period. From this,
we conclude that it is unlikely that the observed light variations are caused by
binary companion.

We subsequently performed a similar analysis as done by \citet{monidvoj1,monidvoj2}
and searched for binarity from the radial velocity data. From the analysis, it
follows that the measurements are perfectly compatible with constant radial
velocities. A Kolmogorov–Smirnov test reveals that the probability of these
results being drawn from a normal distribution (with a dispersion equal to the
observational errors) is equal or higher than about 50\% for each star (namely
48\%, 64\%, and 86\% for vEHB-2, vEHB-3, and vEHB-7).

Going further, we estimated the probability of these stars being
undetected binaries. The most common close companions of extreme horizontal
branch stars are compact objects such as white dwarfs. Hence, we simulated
systems of $0.49+0.49\,M_\odot$ stars (typical of such systems), in circular
orbits (because the short periods suggest a previous common envelope phase, which
circularizes the orbits), with orbital period equal to the photometric one, an
isotropic distribution of the angle of inclination of the orbit (hence uniform
in $\sin i$), and a random phase. We considered a system to be "undetected" if the
observed radial velocities (at the epochs of observations) show a maximum
variation lower or equal to that observed. We found that the probability that
the studies stars hide undetected binary are 1.74\% for vEHB-2, 0.68\% for
vEHB-3, and <0.01\% for vEHB-7. The differences stem from different periods of
variability. In conclusion, the angle of inclination cannot explain the
lack of evidence with respect to binarity.

For the radial velocity analysis, we assumed a canonical mass for the extreme
horizontal branch stars, but the estimated values are much lower. The lower mass
would make the probabilities even lower, because with a smaller mass, the radial
velocity variations would be greater. On the other hand, with lower mass of the companion,
 the binarity could  pass undetected more easily. Consequently, we checked
what the companion mass must be to obtain the probability of an undetected
binary of at least 5\%. This results in 0.27, 0.17, and 0.04$\,M_\odot$ for the
studied stars. As we have already shown, the masses are too low to explain
photometric variations by mean of ellipsoidal variation or reflection effects.
The exception could possibly be vEHB-2, but it has the longest period, which
implies a much larger separation between the components, again arguing against
both tidal and reflection effects.

Another possibility is that the light variations are not due to the star itself,
but due to another star that coincidentally appears at the same location on the
sky. However, it is difficult to find such types of variable stars that would
correspond to observations. Pulsating stars of RR~Lyr type, which are indeed
found on horizontal branch of globular clusters, have significantly shorter
periods \citep[e.g.,][]{skarbla,tessmolnar}. The period of variability better
corresponds to Cepheids. Type II Cepheids may correspond to low-mass stars that
left the horizontal branch \citep{cefvyv} and are indeed found in globular
clusters \citep{braga}. However, they are much brighter in the visual domain than
stars studied here. On the other hand, classical Cepheids corresponds to blue
loops on evolutionary tracks of stars that are more massive than appear in
globular clusters now \citep{nei}. This would imply a  distant background object
that is younger than the cluster. However, taking into account the fact that extreme horizontal
branch stars constitute just a very small fraction of cluster stars, we consider
a chance alignment in three of them to be very unlikely.

\subsection{Temperature spots}

\citet{mombible} pointed out that the observed photometric variations could be
caused by temperature spots. Such spots are predicted to be caused by shallow
subsurface convective zones that may be present in hot stars
\citep{cabra,cabra19} and connected to surface magnetic fields. This
could  indicate the presence of either a \ion{He}{ii} or deeper Fe convective zone. However, helium is
significantly underabundant in studied stars and a corresponding region of
helium underabundance may extend deep into the star \citep{miriri}. As a result, the
\ion{He}{ii} convection zone may be absent \citep{quievy}, as indicated also by
our evolutionary models (Sect.~\ref{kapevol}).
%leaving just Fe convective zone.

The studied variability seems to be stable on a timescale of years, while the
subsurface convection zones were invoked  to explain variability that is more stochastic in nature \citep{cantstoch} and has a significantly lower
amplitude. Subsurface convection was suggested to drive corotating interacting
regions in hot star winds \citep{duo}, which require more spatially coherent
structures, but it is unclear whether they are persistent in the course of
hundreds of days. Based on the analogy with cool star spots and considering
photometric observations of hot stars \citep{chemof,ramxiper,arholeo}, we
consider this possibility to be unlikely. Moreover, the iron convective zone
appears directly beneath the stellar surface, therefore, it does not seem
likely that the magnetic fields can cause large variations of stellar radius
\citep[c.f.,][]{fulmat}.
%~/promhv/esoehbpromhv.aa/mesa/0.5M p1/LOGS_cut/profile2.data 

We have detected variations of the effective temperature
(Figs.~\ref{ehb2prom}--\ref{ehb7prom}), but they predict a greater amplitude of
light variability than what has been observed. To reduce the amplitude, we introduced
additional variations of radius, which cause variability of surface gravity. The
detected variations of surface gravity are in conflict with models of temperature
spots. We tested this by fitting synthetic spectra derived from combination
of spectra with different effective temperatures, but the same surface gravities.
This should mimic the spectra of a star with temperature spot(s). The fit
provided an effective temperature between the temperatures of combined spectra,
but the surface gravity remained nearly constant and equal to the surface
gravities of individual spectra.

About one-third of stars with spots show complex light curves with a
double-wave structure \citep{jagelka}. However, all the light curves observed by
\citet{mombible} are much simpler and consist of just a single wave. This also
is an argument against the notion of spots causing the photometric variability of the studied
stars.

The model of temperature spots can be further observationally tested using
spectropolarimetry, which should be able to detect accompanying weak magnetic
fields. Hot spots could be also detected from radial velocity variations, which
should show a minimum at about a quarter of a phase before the light maximum. This phase
variability is opposite to radial velocity variations due to pulsations, which
show a maximum at a quarter phase before the light maximum. 

\section{Discussion}

\subsection{Stellar masses}

Stellar masses of studied stars derived from spectroscopy and photometry are
rather low for single core-helium burning objects. Although the uncertainties
are typically very large, the masses are systematically lower than a canonical
mass of isolated subdwarfs \citep{heberpreh}. Comparable mass problems also appear
in other studies of globular cluster horizontal branch stars with similar
temperatures \citep{moniomega,sabuvjas,shotomcen2}.

 The cause of this problem is unclear \citep[see the discussion
in][]{monipec}. With a fixed surface gravity from spectroscopy, a higher mass
requires larger radius. This can be achieved either by significantly lower
V~magnitude, higher distance modulus, higher reddening, or lower bolometric
correction. A lower apparent magnitude is unlikely. The Gaia distance modulus of
$\omega$~Cen is slightly lower than the value adopted here \citep{omcengaia}
worsening the problem even more. The adopted reddening agrees with independent estimations
\citep{kalamitaomcen,bonomcen}. The bolometric corrections might be uncertain
and, indeed, \citet{bstar2006}  reported slightly lower value than adopted here.
However, this alone would not solve the problem. Our analysis using the model
atmosphere fluxes computed here, with help of Eq.~(1) from \citet{bstar2006}, shows that
a lower helium abundance slightly increases the bolometric correction, thus
worsening the discrepancy once again.

This leaves the uncertainties of parameter determinations from spectroscopy as the
only remaining cause of overly low derived masses of studied stars connected with the analysis. The true
uncertainties could be higher than the  random errors (given in Table~\ref{hvezpar})
when accounting for systematic errors (Sect.~\ref{ransyer}).

%To better understand the impact of uncertainties of atmospheric analysis, we
%fixed the surface gravity of vEHB-7 to a value by 0.3\,dex higher than
%determined from spectroscopy (Table~\ref{hvezpar}) and repeated the spectral
%analysis. Merely increasing the gravity would lead to a more acceptable value of
%stellar mass $0.35\,M_\odot$, but the spectral analysis gives a higher effective
%temperature of $22\,000\,$K, which in turn gives mass of $0.33\,M_\odot$.

Lower mass subdwarfs may also originate due to some more exotic evolutionary
processes. Subdwarfs with mass lower than the canonical one are found among
field stars, but they typically appear in binaries \citep{kupkompakt} and
require binary interaction for explanation \citep{altmalo}. Moreover, stars with
initial masses of about $2\,M_\odot$ may ignite helium in a non-degenerate core with
mass as low as $0.32\,M_\odot$ \citep{han,aramesa}. However, the lifetime of such stars
is at odds with expected age of $\omega$~Cen. In any case, low-mass white
dwarfs with mass around $0.2\,M_\odot$  were detected, which are considered to be
connected with hot subdwarfs \citep{heberpreh}. A lower mass of about
$0.3\,M_\odot$ was also predicted for blue large-amplitude pulsators in the
context of their He pre-white dwarf nature \citep{takynevyslo,romcoblap}.
However, alternative models for these stars propose either helium shell or core
burning subdwarfs with higher masses \citep{wuli,xislup}.

\begin{figure*}[t]
\includegraphics[width=\textwidth]{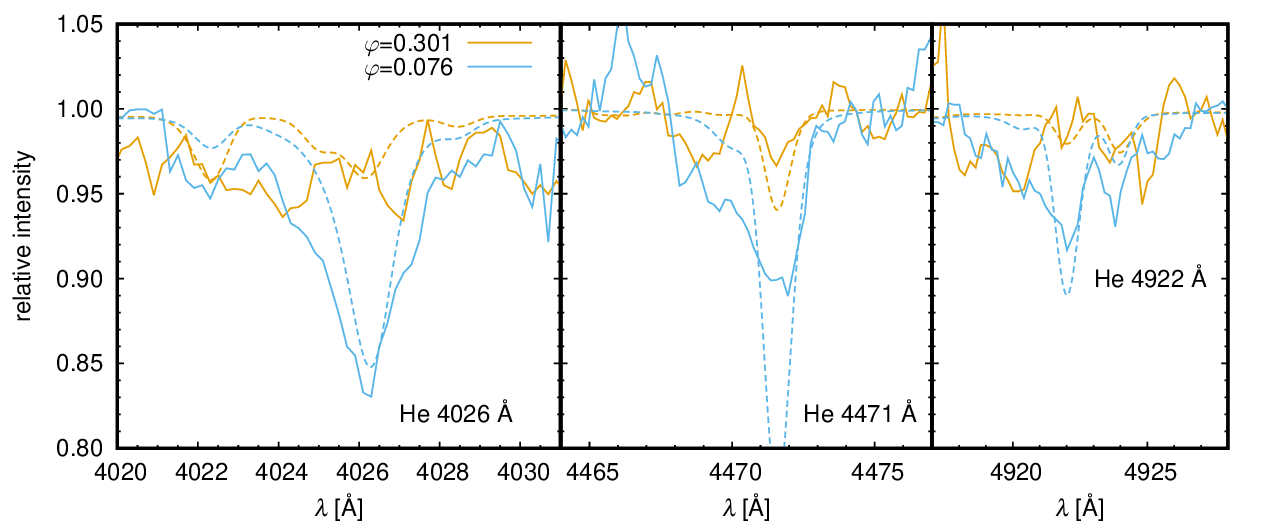}
\caption{Comparison of observed (solid lines) and predicted (dashed lines) 
helium line profiles for two different phases in the spectra of vEHB-3.}
\label{ehb3_he}
\end{figure*}

\begin{figure*}
\centering
\includegraphics[width=0.7\textwidth]{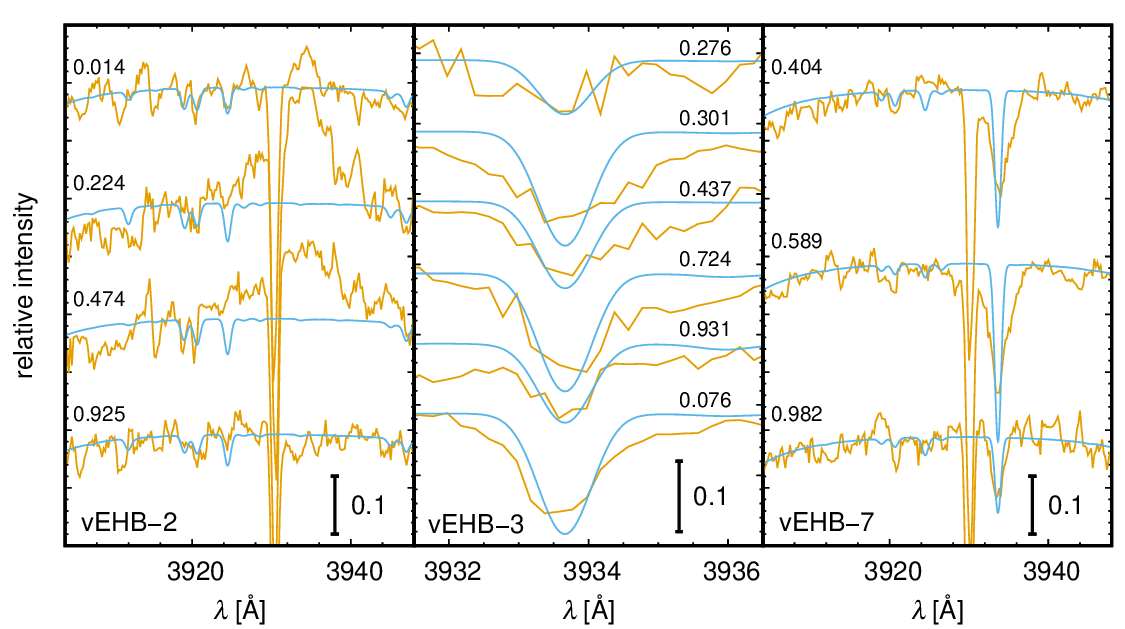}
\caption{Phase variability of \ion{Ca}{ii} 3934\,\AA\ line. Plotted for
individual studied stars for all phases (denoted in the graph). The plot
compares observed spectra (yellow lines) with predicted spectra (blue lines).
The vertical scale denotes fraction of the continuum intensity. The spectra were
shifted to the stellar rest frame.}
\label{ehb1237_ca}
\end{figure*}

\subsection{Tension with parameters from literature}

For star vEHB-7, \citet{shotomcen} determined slightly higher effective
temperature and surface gravity. However, their data were collected by the FORS
spectrograph, which has a lower resolution that X-shooter. We simulated the
consequences of using low resolution spectra for the derived parameters and we smoothed the
data by a Gaussian filter with dispersion of 3\,\AA, which roughly corresponds to
a FORS resolution, according to the user
manual\footnote{https://www.eso.org/sci/facilities/paranal/instruments/fors/doc.html}.
The fitting of spectra with a lower resolution has  systematically provided higher
effective temperatures by about 500\,K and higher surface gravities by about
0.2\,dex. This partially explains the differences in the derived parameters.

Similarly, \citet{momega} found a higher effective temperature for vEHB-2.
However, these authors used spectra with shorter interval of wavelengths. Our tests have
shown that this can lead to differences in the effective temperature of about
1000\,K and surface gravity of about 0.1\,dex. This could be one of the reasons
behind the differences in the determined parameters.

The effective temperature and surface gravity were derived from the fits of models
with underabundances of heavier elements, although we do see that iron shows an overabundance with
respect to the solar value (Table~\ref{hvezpar}). \citet{momocfe} alleviated this
problem by using models with higher abundances of iron. However, the comparison
of spectra from the BSTAR2006 grid \citep{bstar2006}, with different iron abundances,
showed nearly identical hydrogen line profiles. Therefore, we conclude that this
is not a significant problem for the parameter determination presented here.

Two of the variable horizontal branch stars detected by \citet{mombible} in
NGC~6752 were subsequently analyzed by \citet{shotomcen2}. It turned out than only one of
them is a genuine horizontal branch star, while the other was instead
classified as a blue straggler. The horizontal branch star has very similar
atmospheric parameters as obtained here and it also has a slightly lower mass
than typical for horizontal branch stars \citep[Fig.~14,][]{shotomcen2}, albeit
higher than that derived here.

\subsection{Line variability}
\label{kapcarprom}

We detected variability among the helium and calcium lines, which is also likely to be phased
with the variability period (Figs.~\ref{ehb3_he} and \ref{ehb1237_ca}). Such
variability may indicate presence of spots. However, our tests have shown that the
abundances are too low to cause any significant light variability
(Sect.~\ref{kapabskvrn}). Classical chemically peculiar stars may show vertical
abundance gradients in the atmosphere \citep[e.g.,][]{leblahb,vesel}, but this
would not help to explain the light variability because the opacity in the
continuum-forming region is decisive. Moreover, the line profiles are unusually
broad in some cases and the calcium line may even appear in the emission. This is the case
for the star vEHB-2 (Fig.~\ref{ehb1237_ca}). In addition, emission is also likely to appear in
one spectrum of vEHB-7, which was not included in the present analysis due to
its low S/N.

The unusual variability of these lines and the appearance of emission could be
perhaps connected with shocks that propagate throughout the stellar atmosphere
as a result of pulsational motion \citep{cernypulz,jefhydropul}. The shock may
possibly heat the atmosphere and induce the emission in the \ion{Ca}{ii} 3934\,\AA\
line. The shock appears around the phase of minimum radius (maximum gravity),
which agrees with spectroscopy of vEHB-2 (Fig.~\ref{ehb1237_ca}).

\subsection{Evolutionary considerations}
\label{kapevol}

To better constrain the nature of the light variability of the studied stars, we
simulated their internal structure using the MESA code\footnote{We used MESA
version 22.11.1.}
%\citep{Paxton2011, Paxton2013, Paxton2015, Paxton2018,
\citep{Paxton2019, Jermyn2023}. We selected a model star with an initial mass of
$2.2\,M_\odot$, which starts to burn helium at the moment when the core mass is
close to the mass of the stars used in this study \citep{han}.

We simulated the evolution of a star from the pre-main sequence until the
initiation of helium-burning in the core. By setting the mass fraction of heavy
elements to $Z=0.0006$ and incorporating convective premixing and the Ledoux
criterion\footnote{For reference see
https://docs.mesastar.org/en/latest/index.html.}, we ensured a similar
representation of the stellar conditions. Compared to standard models, we also
included silicon and iron elements to account for the essential constituents
found from observations. Afterward, we stripped the star's envelope, leaving
behind only the helium core enveloped by a hydrogen-rich outer layer with mass
of $0.01\,M_\odot$. This process allowed us to imitate the physical structure found
in horizontal branch stars. We also evolved a similar model star with an
additional accreted mass of $0.001\,M_\odot$  mirroring the composition of the
surface material deduced in vEHB-2.

Our approach is similar to the work of \citet{han} and gives comparable
effective temperatures ($25-30\,$kK) and surface gravities ($\log g \approx5.5$)
during the helium-burning phase. Contrary to \citet{han}, who were able to
create the lowest mass helium-burning star with a zero-age main sequence mass of
$1.8\,M_\odot$ for $Z=0.004$, we found that our models did not allow us to use such a low initial
mass. This suggests that compactness of the inner core was greatly affected by
including the heavy elements,  thereby creating helium or hydrogen flashes
for lower initial masses.

We noticed a notable disparity between the non-accreted and accreted models. While
the models with near solar helium fraction ($Y=0.24$) displayed a convection
layer near the surface, the layer disappeared after the accretion of helium-poor
material. Therefore, models do not predict any subsurface convective region for
a chemical composition derived from observations.

Alternatively, the parameters of the stars correspond to stars in the post-red
giant evolutionary state \citep{hall8000}. In that case, the variability of
studied stars could be connected with instability of hydrogen-burning on the
surface of a degenerate core \citep{shebi}, which could lead to periodic
behavior \citep{perzable}.

\subsection{Random and systematic errors}
\label{ransyer}

Random errors among the parameters in individual phases were determined using the
Monte Carlo method. However, there might be certain errors in the analysis that
could not be described by random errors. To better assess the statistical
significance of the results, we searched the ESO X-shooter archive for multiple
observations of subdwarfs. We focused on subdwarfs listed in the catalog from
\citet{subkatii}, which  have similar parameters to those of the horizontal branch stars studied
here.

We selected the field hot subdwarf EC\,01510-3919, which has four spectra from
two nights available in total. We analyzed the spectra in the same way as we
did for horizontal branch stars. The analysis provided
$T_\text{eff}=20\,440\pm90\,\text{K}$ and $\log g=4.73\pm0.02$, in a good
agreement with parameters determined by \citet{orisek}.

The maximum differences between effective temperature and surface gravity
estimates from individual spectra were about $200\,$K and $0.03\,$dex,
respectively. Although the S/N ratio of the spectra is roughly a factor of two
higher than for globular cluster stars, this further demonstrates that the detected
variations of the effective temperature and surface gravity are likely to be real.
Moreover, the analysis also shows that the mismatch between observed and fitted
variations of surface gravity of vEHB-7 could be of a random origin.

We studied the effect of continuum normalization on the uncertainty of
parameters. To test the influence of normalization, we multiplied the absolute
data by a smooth function and repeated the analysis again (including
normalization). This had a small effect on the derived parameters. We performed
additional tests by restricting the number of lines used for the analysis. This
also led to similar variations as those we detected, albeit with a larger scatter.

Unlike the random errors considered here, the systematic errors are much more
difficult to estimate. They may be connected with uncertainties of parameters
such as oscillator strengths, NLTE model ions, continuum placement, and
selection of lines for the analysis \citep{statchyb}. The systematic errors
can be roughly estimated from a comparison of derived parameters with
independent estimates from the literature, which gives an error of about $1000\,$K
in the effective temperature, and 0.1\,dex in the surface gravity and abundances.
However,
{unlike the random
errors, the systematic errors affect all the measurements in
approximately the same way. Therefore,}
because this study is focused mainly  on the  origin of the light
variability  connected to differences among individual spectra, the
{systematic} errors are of a lesser importance.

\section{Conclusions}

We analyzed the phase-resolved spectroscopy of three periodically variable
extreme horizontal branch stars from the globular cluster $\omega$~Cen that were detected
by \citet{mombible}. We determined the effective temperatures, surface
gravities, and abundances in individual photometric phases.

We detected the phase variability of the apparent effective temperature and surface
gravity. The effective temperature is the highest during the light maximum. We
did not detect any strong variability of abundances that could explain the
observed photometric variations; neither did we detect any significant radial
velocity variations that could point to the binarity. Instead, the photometric
and spectroscopic variability can be interpreted in terms of pulsations. This is
additionally supported by
%the period-luminosity relationship that the studied stars fulfill and
the  anomalous profiles of helium and calcium lines that point to intricate
atmospheric motions. The effective temperatures of these stars, $21-25\,$kK, and
the surface gravity correspond to extension of PG~1716 stars or
blue, high-gravity, large-amplitude pulsators toward lower
temperatures, albeit with much longer periods. Given the effective temperature
of these stars and the length of their periods, we propose that the pulsation of
these stars are due to g modes initiated by the iron opacity bump. However, the
length of the periods of the
order of day is in strong conflict with Ritter's law.

Surface temperature spots provide the only viable alternative explanation for the
light variability. Nevertheless, the detection of surface gravity variations in
studied stars and the existence of complex line profile variations of the helium and
calcium lines offer additional support for the pulsational model.

The metal-deficient chemical composition of these stars corresponds to the
horizontal branch of globular clusters. One exception is iron, with a roughly solar
chemical composition that is perhaps due to radiative diffusion. On the other hand,
helium has significantly subsolar abundance that is likely due to gravitational
settling.

We estimated the masses of these stars from spectroscopy and photometry in the range
of $0.2-0.3\,M_\odot$. This  value is too low for helium-burning stars, but
similar estimates were obtained previously for horizontal branch stars.

\begin{acknowledgements}
We thank Dr.~Yazan Momany for valuable comments on the paper and Dr.~Petr
Kurf\"urst for the discussion.
Computational resources were provided by the e-INFRA CZ project (ID:90254),
supported by the Ministry of Education, Youth and Sports of the Czech Republic.
\end{acknowledgements}

\bibliographystyle{aa}
\bibliography{papers}

\end{document}